\documentclass[11pt]{article}
\usepackage{psfig,epsf,graphics}
\newcommand \be {\begin{equation}}
\newcommand \ee {\end{equation}}
\newcommand \bea {\begin{eqnarray}}
\newcommand \eea {\end{eqnarray}}

\newcommand \bi {\bibitem}
\newcommand \s {\sigma}

\date{\today}

\begin{document}
\begin{titlepage}
\begin{center}
{\large\bf Dynamical AC study of the critical behavior in Heisenberg spin glasses.}\\[.3in]
  {\bf Marco Picco$^{a}$ and Felix Ritort$^{b}$} \\
  $^{a}$ Laboratoire de Physique Th\'eorique et Hautes Energies
         \footnote{Unit{\'e} Mixte de Recherche CNRS UMR 7589, associ\'ee
         \`a l'Universit\'e  Pierre et Marie Curie, PARIS VI
         et \`a l'Universit\'e  Denis Diderot, PARIS VII.},\\
  Bo\^{\i}te 126, Tour 16, 1$^{\it er}$ {\'e}tage,\\
  4 place Jussieu, F-75252 Paris CEDEX 05, France\\
  $^{b}$ Departament de F\'{\i}sica Fonamental, Facultat de F\'{\i}sica,\\
         Universitat de Barcelona\\ 
         Diagonal 647, 08028 Barcelona (Spain)

\end{center}
\vskip .15in
\centerline{\bf ABSTRACT}
\begin{quotation}

We present some numerical results for the Heisenberg spin-glass model
with Gaussian interactions, in a three dimensional cubic lattice. We
measure the AC susceptibility as a function of temperature and
determine an apparent finite temperature transition which is
compatible with the chiral-glass temperature transition for this
model.  The relaxation time diverges like a power law $\tau\sim
(T-T_c)^{-z\nu}$ with $T_c=0.19(4)$ and $z\nu=5.0(5)$.  Although our
data indicates that the spin-glass transition occurs at the same
temperature as the chiral glass transition, we cannot exclude the
possibility of a chiral-spin coupling scenario for the lowest
frequencies investigated.

\vskip 0.5cm
\noindent



\end{quotation}
\end{titlepage}

\nobreak

Dynamical AC measurements in metallic spin glasses (e.g. CuMn) show the
existence of a cusp in the in-phase component of the AC susceptibility
located at a temperature value $T_m$ that shifts to lower temperatures
as the frequency $\omega$ of the AC field
decreases~\cite{BinYou86,Thol1}. The relaxation time is then given by
the inverse of the AC frequency and increases very fast as the
temperature decreases. Several laws have been proposed to describe this
dependence such as the Vogel-Fulcher law \cite{Thol2} or super-Arrhenius
behavior~\cite{BinYou84}. However, most experimental data \cite{SouTho85} show
that $T_m(\omega)$ can be well fitted to a power law over several orders
of magnitude, $T_m(\omega)-T_c\sim \omega^{\frac{1}{z\nu}}$.  $T_c$ is
the value at which the characteristic relaxation time diverges and the
power law behavior has been interpreted as a signature of a phase
transition at $T_c$.

Most metallic spin-glasses are characterized by low spin anisotropy
where a description of the magnetic moments in terms of continuous
Heisenberg spins seems appropriate.  The vast majority of theoretical
studies in this model (analytical and numerical) have considered the
critical behavior by studying the equilibrium properties. From these
studies it emerges that this model undergoes a chiral-glass phase
transition at a finite temperature \cite{kawamura,hk} while it was for
long time believed that the spin-glass transition occurs only at zero
temperature \cite{OYS}. A chiral-glass transition occurs if the
spin-reflection symmetry is broken but not the spin-rotation
symmetry. In more recent works, it was claimed that at the chiral-glass
transition, there is also a spin-glass transition
\cite{matsubara,ne,mset,ly,by}. AC studies provide a direct method to
investigate the critical behavior and are relevant as most of the
experimental evidence in favour of the spin-glass transition is based on
such type of measurements. In this letter we report on some dynamical AC
simulations concerning the critical behavior in the Heisenberg
spin-glass model in three dimensions. The goal of this work is then to
find an estimate of the spin-glass transition temperature and the value
of the critical exponent $z\nu$. The model is defined as
\be
{\cal H}=-\sum_{<i,j>}\,J_{ij}\vec{S_i}\cdot{\vec{S_j}} -h\sum_{i=1}^N
S_i^z \; ,
\label{eqEA}
\ee
where the index $i$ runs from 1 to $N=L^3$, ${\vec{S_i}}$ is a vector of
unit modulus and the $<i,j>$ corresponds to a pair of nearest neighbors
spins in a finite dimensional lattice with periodic boundary
conditions. The exchange couplings $J_{ij}$ are taken from a Gaussian
distribution with zero average and unit variance.  Monte Carlo
simulations of (\ref{eqEA}) use random updating of the spins with the
Metropolis algorithm. A spin is randomly chosen and its value is changed
as
\be
\vec{\s_i}\to
\vec{\s_i}+\vec{\delta_i}=\vec{\s_i}+\delta\cdot\vec{r_i}\; ,
\label{motion}
\ee
with $\delta$ a finite number ($\delta = 1$) and $\vec{r_i}$ a vector with random
components extracted for a Gaussian distribution of unit variance. The
new spin after the change in (\ref{motion}) is rescaled in such a way
that it remains of unit length.

Our simulations of the Heisenberg model are done in the following way:
An oscillating magnetic field $h(t)=h_0\cos(2\pi\omega t)$ of frequency
$\omega=\frac{1}{P}$, where $P$ is the period, is applied to the system
and the magnetization measured as a function of time
\be
M(t)=M_0\cos(2\pi\omega t+\phi)\; ,
\label{eqMAC}
\ee
with $M_0$ the intensity of the magnetization and $\phi$ the
dephasing between the magnetization and the field. The origin of the
dephasing is dissipation in the system which prevents the magnetization
to follow the oscillations of the magnetic field. From the magnetization
we obtain the in-phase and out-of-phase susceptibilities defined as
\begin{eqnarray}
\chi'=\frac{M_0\cos(\phi)}{h_0}=\frac{2\int_0^PM(t)\cos(2\pi\omega t)dt} 
{h_0}\label{eqchi1}\\
\chi''=\frac{M_0\sin(\phi)}{h_0}=\frac{2\int_0^PM(t)\sin(2\pi\omega t)dt} 
{h_0}~~~~.\label{eqchi2}
\end{eqnarray}
The dephasing $\phi$ measures the rate of dissipation in the system and
is given by
\be
\tan(\phi)=\frac{\chi''}{\chi'} \; .
\label{tanfi}
\ee
The in-phase and out-of-phase susceptibilities are computed by averaging
the right-hand side of (\ref{eqchi1}), (\ref{eqchi2}) over several
periods $P=\frac{1}{\omega}$. We always averaged at least over 10
periods of time, after discarding the first four periods of time. We also took
averages over several realizations of disorder, at least 6 for the
largest sizes that we simulated, $L=40$. 

\begin{figure}
\begin{center}
\rotatebox{270}
{\epsfxsize=10cm\epsffile{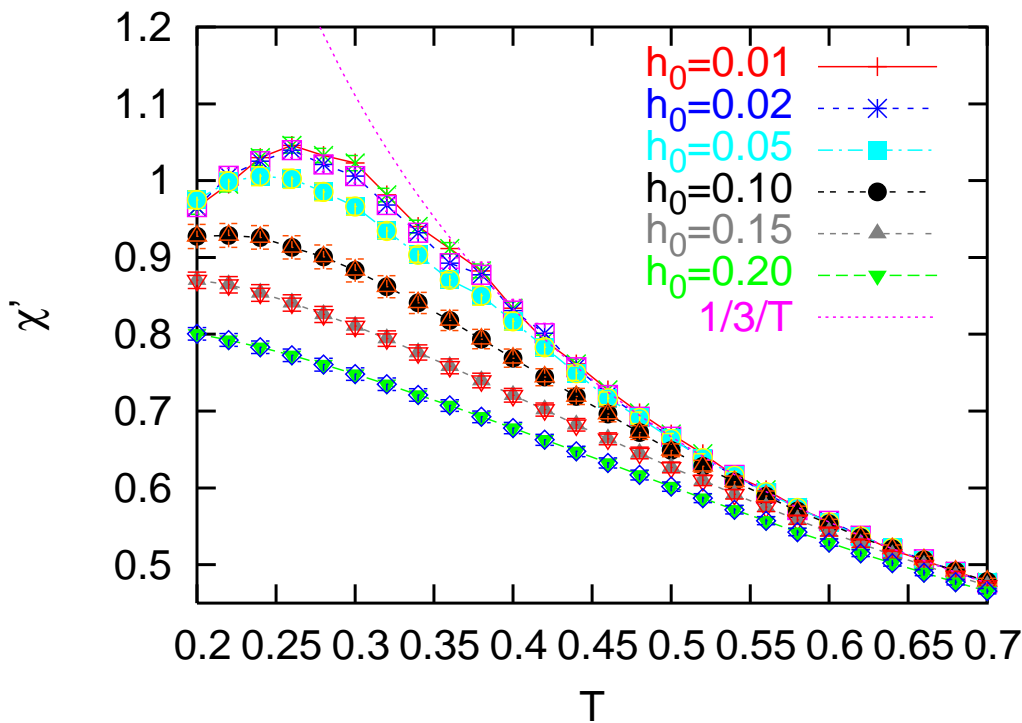}}
\caption{$\chi'$ vs. $T$ for $L=10$, $P=50 000$ and for 
varying value of the amplitude of magnetic field $h_0= 0.02, 0.05, 0.10
, 0.15$ and $0.20$ (from bottom to top). }
\label{fig1}
\end{center}
\end{figure}
For each frequency $\omega$, we determine the
temperature $T_m$ corresponding to the maximum of the in-phase susceptibility
$\chi'$ \cite{VHA,AJM}. This temperature determines the relaxation time $\tau(T_m)=P$.
Alternatively, one could also define $T_m$ as the
value of the temperature at which an inflection point is observed in the
out-of-phase susceptibility $\chi''$.  Next, one can determine the
static transition temperature as well as the critical exponent $z\nu $
using the scaling relation
\be
\tau(T) \simeq  c (T-T_c)^{-z\nu}\; .
\label{relax}
\ee
where $c$ is a constant. In a previous work \cite{PRR}, we had
employed this method to study the Ising spin-glass model and the
Heisenberg spin-glass model. Our findings, for the Ising case, was in
very good agreement with a previous numerical study \cite{AJM,og} as
well as with experimental results \cite{nord1,MatTazMiy89,VHA,nord2}. For the Heisenberg
case, our conclusion was that we had data compatible with a zero
temperature divergence $\tau(T) \simeq T^{-z\nu}$ with a value of
$z\nu \simeq 5.8$. In the present work, we reconsider more carefully
the Heisenberg case. In AC susceptibility measurements there are two
effects that must be carefully evaluated: finite-size effects and the
amplitude of the AC field. Finite-size effects become particularly
important as we move close to the critical temperature where the
correlation length diverges. Since one has no direct access to the
correlation length, one must be sure that the size considered is large
enough. The amplitude of the magnetic field $h_0$ is also important as
we want to keep our measurements in the linear response
regime. Already for the Ising model \cite{AJM}, it was observed that a
too large value of $h_0$ can affect the measurements. But in the case
of the Ising spin-glass model, it is only for rather large values of
the magnetic field $h_0$ that deviations were observed in simulations,
typically for $h_0 \simeq 0.4$.  Here on the contrary, we will see
that one needs to adjust the value of $h_0$ as a function of the
frequency. The lower the frequency, then the lower the temperature
that we want to probe, and the lowest the magnetic field must be. For
the lowest frequency that we simulated, $\omega=1/ 150 000 $, we need
to reduce the amplitude of the magnetic field to $h_0=0.01$. In
Fig. \ref{fig1} we show susceptibility values obtained for $L=10$ and
$P=1/\omega=50 000$, that is a rather small frequency.  We can
see how the position of the maximum is strongly
affected by the value of $h_0$. It is only for values of $h_0 \leq
0.02$ that the maximum converges to $T \simeq 0.26$. However the
largest value of $h_0$ up to which we can locate the maximum of $\chi'$
for a given $P$ does not seem to depend much on the size
considered. Most of the simulation time has been spent at determining
$h_0$ for each frequency. This was done by repeating, for each
frequency, the measurements as shown in Fig.\ref{fig1} for decreasing
values of $h_0$. The values of $h_0$ that we obtained are $0.05$ for
$P \leq 5000$, $0.02$ for $P=15 000, 50 000$ and $0.01$ for $P=150
000$.  

In practice, the fact that one needs to reduce the amplitude of the
applied AC field has an important consequence. Indeed, fluctuations in
the measured susceptibility increase very fast as we decrease $h_0$,
as could be expected from (\ref{eqchi1},\ref{eqchi2}). Thus, to reduce
the errors on the value of $\chi',\chi''$, one needs to increase the
number of simulated samples. Consequently, since $h_0$ has to be
reduced as one increases/decreases the value of the period/frequency,
one must simulate a steadily larger number of samples as the size of
the system increases.
\begin{figure}
\begin{center}
\rotatebox{270}
{\epsfxsize=10cm\epsffile{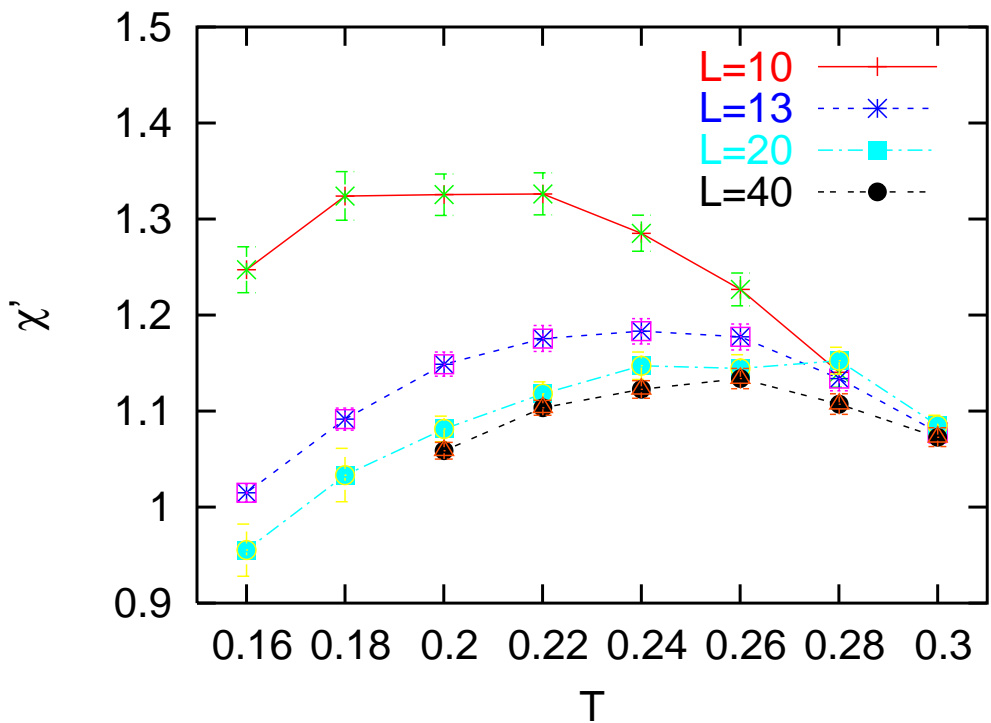}}
\caption{In-phase susceptibility as function of temperature, 
for $h=0.01$, $P=150 000$ and for $L=10,13,20$ and $40$ 
(from top to bottom).}
\label{fig2}
\end{center}
\end{figure}
In Fig. \ref{fig2} we show $\chi'$ as a function of the temperature for
$P=150 000$, $h_0=0.01$ and for different sizes at the lowest frequency
$\omega=1/P=1/150 000$. It emerges from Fig. \ref{fig2} that finite-size
effects are very important and strongly influence the position of the
maximum of the susceptibility. For $L=10$, the maximum is located at
$\simeq 0.20$, while for $L=13$ it has moved to $\simeq 0.24$ and then
it stabilizes close to $\simeq 0.26$ for $L=20$, apparently not changing
anymore for larger sizes. Thus it seems that finite-size corrections can
be very important, meaning that the correlation length must be
rather large. Although the same type of behavior is also observed at other
frequencies, it becomes more evident as we move to lower frequencies. This
might have been expected, as the correlation length
increases when the temperature decreases.  In practice, for all the
simulations that we have performed, we always got the same
susceptibility for $L=20$ and $L=40$, thus we expect that our
measurements for these sizes are free of finite-size effects.

\begin{figure}
\begin{center}
\rotatebox{270}
{\epsfxsize=10cm\epsffile{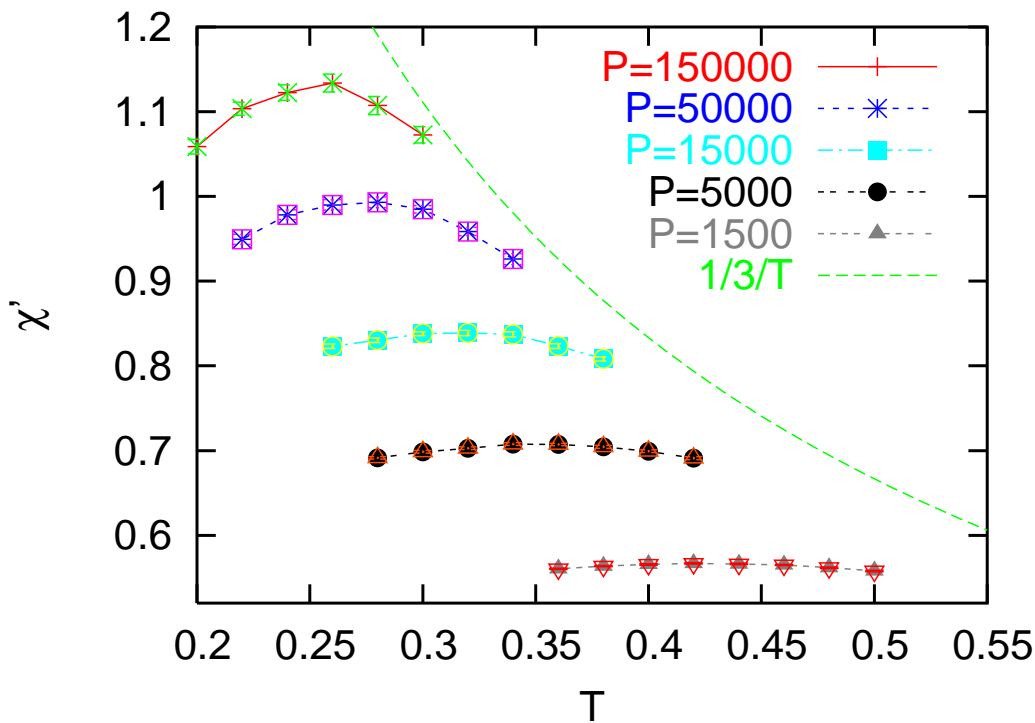}}
\caption{$\chi'$ vs. $T$ for $P=1500, 5000, 15000, 50000$ and $150000$
for $L=40$.}
\label{fig3}
\end{center}
\end{figure}

\begin{figure}
\begin{center}
\rotatebox{270}
{\epsfxsize=10cm\epsffile{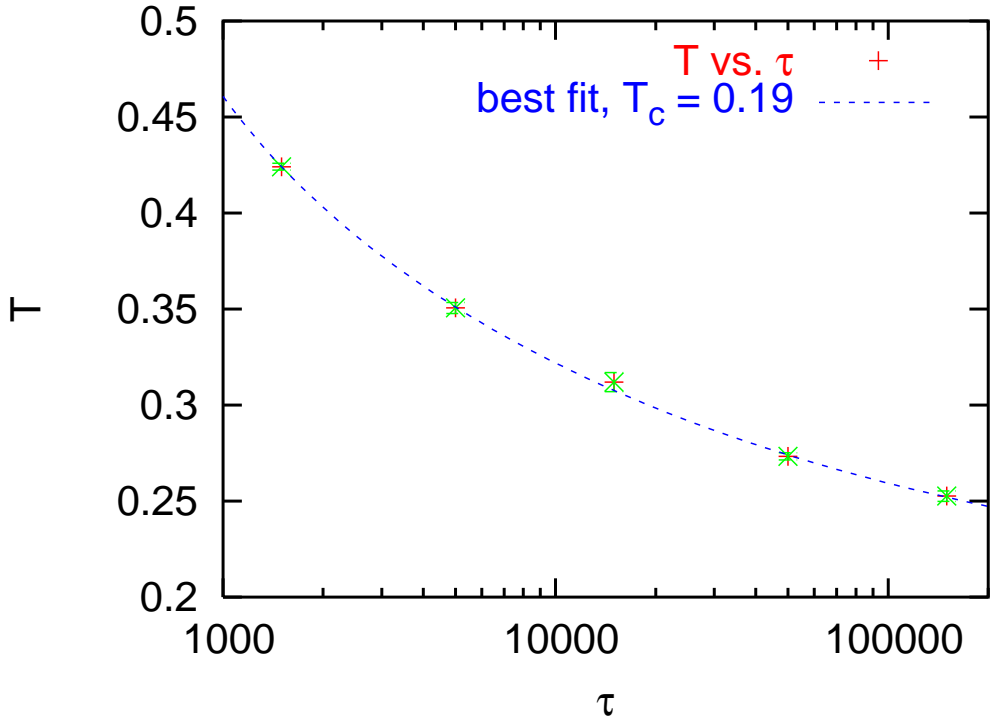}}
\caption{Temperature of the maximum of the in-phase susceptibility
 vs. the period $P$, for $L=40$. The discontinuous line is a best fit
 with (\ref{relax}).}
\label{fig4}
\end{center}
\end{figure}

We will now present our main results for the critical behavior. We have
computed the AC susceptibility for three sizes, $L=10, 20, 40$ and for
$P=1500, 5000, 15000, 50000$ and $150000$. For each of these sizes and
periods, we have adjusted the value of the amplitude of the magnetic
field $h_0$ in order to not see any shift of the maximum of $\chi'$, as
explained above. Also, we have checked that for each value of $P$, the
susceptibility does not change between the sizes $L=20$ and $L=40$. The
data that we obtained for $L=40$ is shown in Fig.\ref{fig3}. We also see
how the value of $T_m$ decreases as $\omega$ decreases. $T_m$ has been
determined as the temperature for which there is a maximum in the
susceptibility after fitting data to a parabolic form.  In
Fig.\ref{fig4}, we show the temperature $T_m$ versus the period
$P$. Data can be fitted to a power law (\ref{relax}) with $T_c=0.19
\pm 0.04$ showing the existence of a finite-temperature
transition in the zero-frequency limit. Since this value is very close
to the one obtained for the chiral glass transition \cite{hk} and the
spin-glass transition \cite{ly}, we have repeated a fit but imposing the
value for $T_c$ obtained in these works, i.e. $T_c= 0.16$, in order to reduce the
number of parameters of the fit. With this condition, we obtain a value of $z\nu =
5 \pm 0.5$ in good agreement with other estimates \cite{hk}.

In this paper we have performed AC susceptibility simulations for the
Heisenberg spin-glass model. By carefully adjusting the value of the
amplitude of the applied magnetic field $h_0$ as a function of the
frequency considered, we have determined the relaxation time
associated to each temperature. Extrapolating the relaxation time to
the limit of zero frequency, we extract a value of critical
temperature $T_g = 0.19 \pm 0.04$ compatible with either the values
obtained for the chiral-glass transition \cite{hk} or the spin-glass
transition \cite{matsubara,ne,mset,ly}. We also computed the exponent
$z\nu = 5.0 \pm 0.5$ in good agreement with previous estimates
\cite{hk} and experimental data for Heisenberg-like models
\cite{SouTho85,VHA,dupuis}. Thus the simplest conclusion is that we
are just observing a spin-glass transition at the same temperature as
the chiral glass transition, in agreement with other recent studies
\cite{matsubara,ne,mset,ly,by}. But this is not the only
possibility. Indeed, in the coupling/decoupling spin-chirality
scenario of Kawamura \cite{kawamura}, one expects that the chirality
will couple at small distances with the spins.  Thus, as far as one
considers only small distances, one can observe physical phenomena
expressed in term of spins while the transition is really on the
chiral parameter. It is only for distances larger than some scale
$L^{*}$ that spins and chirality decouple. This crossover length is
related to the value of the period or frequency required to probe
lengths of order $L^*$ in simulations. The value of $P$ is expected to
be around $10^5 - 10^6$ \cite{k} which is of the same order as the
largest time that we probe in our simulation, $P=150 000$. Thus we
cannot, with the present simulations, exclude the possibility of a
chiral-spin coupling scenario. Simulations for much lower frequencies
are needed in order to probe this coupling/decoupling scenario and
tell if indeed, the spin transition that we observe is due to the
chiral glass transition or not.

{\bf Acknowledgements}
We want to thank F.~Ricci-Tersenghi for discussions in the early
stages of this work. One of the authors (MP) also wants
to thank H.~Kawamura for many helpful discussions and for the
hospitality in Osaka University while part of this work was done.
F. R has been supported by the Spanish grant BFM2001-3525, the STIPCO
EEC network and the SPHINX program of the ESF.


\begin{thebibliography}{99}

\bi{BinYou86} K. Binder and A. P. Young, Rev. Mod. Phys. {\bf 58}, 801 (1986).

\bi{Thol1} J. L. Tholence, Physica B {\bf 126}, 157 (1984).

\bi{Thol2} J. L. Tholence, Physica B {\bf 108}, 1287 (1981).

\bi{BinYou84} K. Binder and A. P. Young, Phys. Rev. B {\bf 29}, 2864 (1984).

\bibitem{SouTho85} J. Souletie and J. L. Tholence, Phys. Rev. B {\bf
32}, 516 (1985).

\bi{kawamura} H.~Kawamura, Phys.~Rev.~Lett.~{\bf 68}, 3785 (1992);
Phys.~Rev.~Lett.~{\bf 80}, 5421 (1998).

\bi{hk} K. Hukushima and H. Kawamura, Phys. Rev. E {\bf 61}, R1008 (2000).

\bi{OYS}  J. A. Olive, A. P. Young and D. Sherrington, Phys. Rev.  B {\bf 9}, 6341 (1986).

\bi{matsubara} F. Matsubara, T.~Shirakura and S.~Endoh, Phys. Rev. B
 {\bf 64}, 092412 (2001). 

\bi{ne} T.~Nakamura and S.~Endoh, J. Phys. Soc. Jpn. {\bf 64}, 2113 (2002).

\bi{mset} F. Matsubara, T.~Shirakura, S.~Endoh and S. Takahashi,
 J. Phys, {\bf A 36}, 10881 (2003).

\bi{ly} L. W. Lee and A. P. Young, Phys. Rev. Lett. {\bf 90}, 227203
(2003), Preprint {\bf cond-mat/0302371}.

\bi{by} L.~Berthier and A.~P.~Young,  Preprint {\bf cond-mat/0312327}.

\bi{VHA} E.~Vincent, J.~Hamman and M.~Alba, Solid State Commun. {\bf 58}, 57 (1986).

\bi{AJM} J.-O.~Andersson, T.~Jonsson and J.~Mattson,
Phys. Rev. B {\bf 54}, 9912 (1996).

\bi{og}  A.~T.~Ogielski, Phys. Rev. B~{\bf 32}, 7384 (1985).

\bi{nord1} N. Bontemps, J. Rajchenbach, R. V. Chamberlin and R. Orbach,
Phys. Rev. B {\bf 30}, 6514 (1984) 

\bi{MatTazMiy89} F. Matsukura, Y. Tazuke and T. Miyadai, J. Phys. Soc. Jpn. {\bf 58}, 3355 (1989).

\bi{nord2} J.~Mattsson, T.~Jonsson, P.~Nordblad, H.~Aruga Katori and
A. Ito, Phys. Rev. Lett.~{\bf 74}, 4305 (1995).


\bi{PRR} M.~Picco, F.~Ricci-Tersenghi and F.~Ritort, Preprint {\bf
cond-mat/0005541 v1}, unpublished.


\bi{dupuis} V.~Dupuis, E.~Vincent, J.-P.~Bouchaud, J.~Hammann, A.~Ito and H. Aruga Katori, 
Phys. Rev.~B {\bf 64}, 174204 (2001), Preprint {\bf cond-mat/0104399}. 

\bi{k} K. Hukushima and H. Kawamura, to be published. 

\end{thebibliography}
\end{document}